# Prior-enhanced reflection spectra imaging with a white light interfermetry

CHENG CHEN, [1] LIXUAN XU,[2] RONG SU, [1*]

[1] Shanghai Institute of Optics and Fine Mechanics, Chinese Academy of Sciences, Shanghai, 201800, China
[2] School of Microelectronics, Shanghai University, Shanghai 201800, China
*Corresponding author: surong@siom.ac.cn

**We present a simple but effective method to extract the reflectivity spectra from the interference signals with a large NA white light interferometry. Numerical simulations and experiments have demonstrated the effectiveness of our method. Furthermore, some insights are also disclosed. © 2020 Optical Society of America**

White light interferometry (WLI) can achieve surface roughness and 3D structural measurement at the microscopic scale. Its full-field imaging capability enables the assessment of surface topography and roughness with both high lateral resolution and sub-nanometer axial noise levels across large fields of view [1-3].

A typical WLI setup includes a broadband light source, Köhler illumination optics, an interference objective, and a mechanical scanner [1]. The Köhler illumination optics projects the light source into the interference objective's pupil. The mechanical scanner provides a smooth continuous scan of the objective in the z-direction. During the scan, a sequence of interference fringes is recorded at each scanning positions. Techniques, such as envelope detection, Fourier domain analysis and correlation between the measured and reference fringes, are typically used to recover topographic information [4-5].

Recent advancements in hardware and image processing techniques have extended WLI capabilities to include reflectivity spectra imaging. J. L. Beverage et. al [6] proposed to use a multi-color-channel camera or a multi-color LED light source to derive colorimetric characteristics of the sample. But spectral resolution is limited by the hardware such as the multi-color-channel camera or the multi-color light source, and there is a trading off WLI's standard performance in exchange for the added colorimetric capability.

Moreover, quantitative reconstruction of reflectivity spectra can also be achieved through Fourier analysis of WLI fringes. Xu et. al [7], and Arnaud Dubois et .al [8] have applied the one-dimensional (1-D) Fourier transform to pixel-wise fringe signals to attain high spectroscopic resolution reflection or absorption spectra in optical coherence tomography (OCT), which has the same basic functional principles as WLI. However, R. Claveau et. al [9] and P. Lehmann et. al [10], have demonstrated that the Fourier magnitude spectrum deviates a lot from the light source's spectrum, when measuring a flat mirror of constant reflectivity across different wavelengths, with a WLI having a numerical aperture (NA) over 0.3. These discrepancies have been confirmed through both simulations and experiments with a statement [10]"the light source is a blue LED with a center wavelength of 460 nm, this results in a maximum effective wavelength of 644 nm for NA=0.55 and 1055 nm for NA=0.9".

R. Claveau and his colleagues have interpreted the Fourier magnitude of a pixel-wise signal as the product of the spectral transfer function of the system and the square root of the sample's reflectivity spectrum in a high NA WLI, while the spectral transfer function of the WLI is calibrated using the Fourier magnitude calculated from the measured fringe signals from a reference sample that has a known reflectivity spectrum. Moreover, a factor was used to characterize the impact of NA on the fringe spacing [11-13], to correct the reconstructed spectra. With this method, the parallel mapping of local spectral and topographic information of gold on an aluminum sample and colored stainless steel have been achieved [14-15]. However, the sensitivity of the Fourier magnitude to measurement noise necessitates lengthy fringe signal acquisition times, often requiring averaging over time and spatial domains to mitigate noise impacts.

In this letter, we report a prior-enhanced spectra inverse algorithm that incorporates a total variation (TV) prior term, to reconstruct the reflectivity spectra from the fringe signal in a high NA WLI. Initially, we establish a linear algebraic representation between the fringe signal and the reflectivity spectra. Subsequently, we solve this ill-posed equation with a TV constraint to the reconstruction process. Detailed procedures of the algorithm will be outlined.

In general, the 1-D Fourier analysis of the pixel-wise WLI signal starts with NA equal 0. Here, we use the 1-D signal model that takes the NA effect into consideration. The WLI signal at pixel-level is an integral over all points in the pupil plane and over all wavelengths for the ray bundle contributions [16],

$$I(z) = \int_0^\infty \int_{\psi_1}^{\psi_2} g(k,\psi,z) P(\psi) s(k) sin(\psi) cos(\psi) d\psi dk, \quad (1)$$

where $z$ denotes the scanning position, $k = 1/\lambda$ is the wave number at an illumination wavelength $\lambda$, $\psi$ is the incident angle within the range of $[\psi_1, \psi_2]$, $s(k)$ denotes optical spectrum distribution of the light source, $P(\psi)$ is the intensity distribution in the pupil plane of the objective and $g(k,\psi,z)$ denotes the interference signal for a single ray bundle at incident angle $\psi$ [17],

$$g(k,\psi,z) = 2\mathrm{Re}\{r_r(k,\psi)^* r_s(k,\psi)\exp[j4\pi k(h_s - z)\cos\psi + j\Delta(k,\psi)]\}, \quad (2)$$

where $r_r$ denotes the reference reflection coefficient, including both the beam splitter and the reference mirror, $r_s$ denotes the sample reflection coefficient, including, e.g., the transmission of the beam splitter, $h_s$ is the sample surface height, and $\Delta$ is an adding phase term associated with chromatic dispersion of the WLI system. The background signal is not taken into consideration here.

Here, we neglect the spectral dependence on the incident angles. Before delving deeper, we introduce a notation in linear algebra. Typically, a '2-D matrix' in bold, italic, uppercase letters; a '1-D vector' in bold, italic, lowercase letters; and a scalar value in italic, lowercase letters. In a discrete form, the spectral distribution of a light source can be discretized at $n$ equally spaced sampling wavenumbers, denoted as $k_1, k_2, \ldots, k_n$, within the spectral bandwidth. Thus, the discrete signal in Eq. (1) can be simplified as

$$\boldsymbol{y} = \boldsymbol{A}\boldsymbol{r}, \quad (3)$$

where $\boldsymbol{y}$ is the $m$ point signal vector, $\boldsymbol{A}$ is the measurement matrix, and $\boldsymbol{r} = |\boldsymbol{r}_r \odot \boldsymbol{r}_s \odot \boldsymbol{s}|$ is the real-valued spectral vector. And $\boldsymbol{r}_r, \boldsymbol{r}_s$ and $\boldsymbol{s}$ denote the discrete vector of complex reflection coefficient of the reference part, the vector of complex reflection coefficient of the sample, and the vector of spectral distribution of light source, respectively. Here, $\odot$ is the element-wise multiplication operation, and $\boldsymbol{r}_s$ is assumed to be 1 for simplicity. The ($i^{\text{th}}, j^{\text{th}}$) entry $a_{ij}$ in the measurement matrix $\boldsymbol{A}$ is given as

$$a_{ij} = 2\int_{\psi_1}^{\psi_2} \cos[4\pi k_j(h_s - z_i)\cos(\psi) + \Delta + \Delta_s]P(\psi)\sin(\psi)\cos(\psi)d\psi, \quad (4)$$

where subscript $i$ denotes the $i^{\text{th}}$ scanning position, subscript $j$ denotes the $j^{\text{th}}$ wavenumber, and $\Delta_s$ denotes phase difference between the complex reflection coefficients between the reference part and the sample's part.

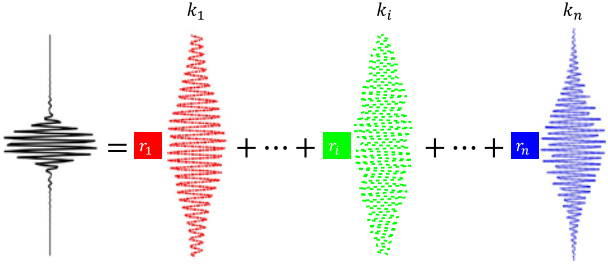

Fig. 1. Linear interpretation of the WLI signal.

The linear representation of the WLI signal in Eq. (3) straightforwardly derives from the interpretating the WLI signal as the sum of a series of monochromatic WLI signals at discrete wavenumbers $k_1, k_2, \ldots, k_n$, each weighted by the corresponding values from the spectral vector $\boldsymbol{r}$, as shown in Fig. 1. And each column in measurement matrix $\boldsymbol{A}$ is denoted as a monochromatic WLI signal vector. This not only lays the foundation of our algorithm, but also offers a computationally efficient method - approximately two-orders of magnitude faster than the direct integration - to simulate the WLI signal.

In an ideal situation, without camera noise, vibrations, or scanning errors, Equation (3) could be readily solved using the least squares method to obtain the spectral vector $\boldsymbol{r}$. With perfect knowledge of the spectral distribution of the light source and the spectral response of the monochrome camera, sample's absolute reflection coefficient can be accurately obtained. However, Eq. (3) is an ill-posed problem, as the solution for $\boldsymbol{r}$ is highly sensitive to small perturbations in both the measurement matrix $\boldsymbol{A}$ and the measured values $\boldsymbol{y}$. The inherent nature of Eq. (3) can be understood from different perspectives. In the spatial domain, monochromatic fringe signals, as shown in Fig. 1, are partially correlated at adjacent wavenumbers. This correlation indicates that the columns in measurement matrix $\boldsymbol{A}$ are not orthogonal to each other, which can lead to a large condition number for matrix $\boldsymbol{A}$, particularly in a high NA WLI. In the spatial frequency domain, the 1-D Fourier transform of the monochromatic fringe signal results in the 1-D transfer function (TF) at each wavenumber. And it is straightforward that the support of TFs at adjacent wavenumbers overlaps in nonzero NA cases, as the monochromatic TF covers a substantial range in Fourier domain [16]. As the NA increases, the support of the monochromatic TF spans a larger range in the Fourier domain, exacerbating the overlapping problem and further degrading the condition of the measurement matrix. At zero NA, the spatial frequency components are separate at different wavenumbers, and Fourier analysis of the pixel-level fringe signal can precisely yield the output the spectral vector, which underpins the basic principle of the Fourier Transform Infrared Spectroscopy.

Furthermore, the spectral distribution of the light source $\boldsymbol{s}$ may exhibit low intensities at certain wavenumbers, which can also deteriorate the reconstruction performance of sample's absolute reflection coefficient. If the reflection coefficient varies slowly along the wavenumber, the stability of Eq. (3) could be improved by modifying the structures of the measurement matrix $\boldsymbol{A}$. We can divide the $n$ wavenumbers into $n_c$ wavenumber channels, where each channel encompasses the combined effect from $n/n_c$ wavenumbers. As shown in Fig. 2, the Eq. (3) can be reformulated as

$$\boldsymbol{y} = \boldsymbol{A}_c \boldsymbol{r}_c, \quad (5)$$

where the conditioned measurement matrix $\boldsymbol{A}_c$ has $n_c$ columns, and each column is the sum of monochromatic fringe signals within the corresponding wavenumber channel with the spectral values of the light source into consideration, and $\boldsymbol{r}_c$ denotes the sample's absolute reflection coefficient at the center wavenumbers at $n_c$ channels.

The conditioned operation not only mitigates the influence of the light source's low-intensity regions on the sample's reflection coefficient reconstruction, but also offers the flexibility to adjust the number of wavenumber channels based on practical measurement requirements. For example, in cases where the NA does not exceed 0.3, the condition number of the conditioned measurement matrix is significantly lower compared to higher NAs, such as 0.55. Therefore, the number of wavenumber channels can be increased to enhance spectral resolution. Conversely, if the measured WLI signals exhibit high noise levels, reducing the number of wavenumber channels can make the reconstruction more robust to noise.

Although the conditioned operation is effective in reducing the condition number of the measurement matrix, other techniques are still needed due to numerous errors sources that cause errors to the measurement matrix and the measured signals. To further enhance the stability of the reconstruction, we would borrow the well-recognized regularization methods in hyperspectral imaging field [18] to our reflectivity or reflection spectrum reconstruction application. This regularization term could be interpreted as the prior information about the desired reflection spectrum, transforming the conventional maximum likelihood estimation (e.g., least squares method) into the maximum a posteriori optimization problem.

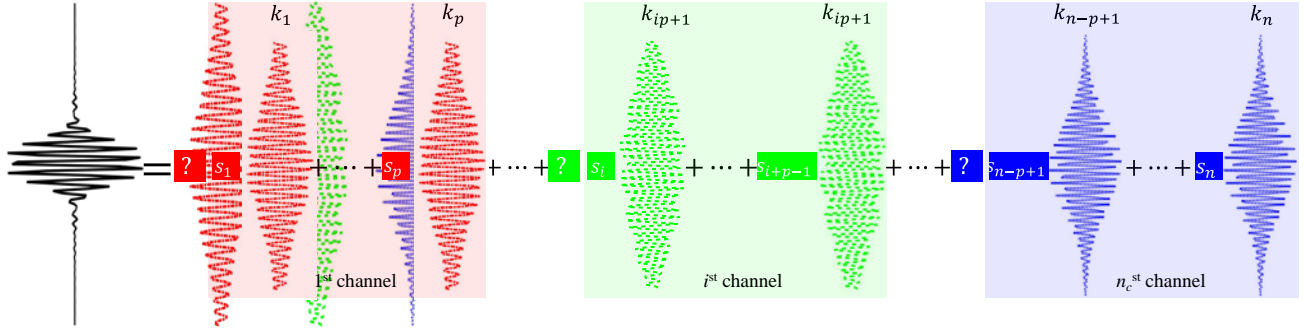

Fig. 2. Linear conditioned representation of a WLI signal.

We leverage the ℓ-1 regularization of the absolute reflection coefficient in a designed space, and assume that the WLI signal is polluted by independently identically distribution (i. i. d.,) Gaussian noise, then the maximum a posteriori framework for Eq. (5) can be given as

$$r_c = \mathrm{argmin}_x \left\{ \tfrac{1}{2} |y - A_c x|^2 + \tau \Phi(x) \right\}, \quad (6)$$

where $\tfrac{1}{2}|y - A_c x|^2$ is the fidelity term that measures the discrepancy between the measured data and the model prediction, $\Phi(x)$ represents the sparsity-promoting penalty on the reflection coefficient spectrum, and $\tau$ is a regularization parameter that balances these two aspects. This approach not only mitigates the influence of noise and other error sources but also ensures that the solution captures the essential characteristics of the reflection coefficient spectrum with reduced complexity. Here, we choose the simple total variation regularization term [18].

The Eq. (6) can be solved with the well-known alternating direction method of multipliers (ADMM) method [19] by splitting the optimization problem into three subproblems as following,

$$\begin{aligned} x^{t+1} &= \mathrm{argmin}_x \left\{ \tfrac{1}{2}|y - A_c x|^2 + \tfrac{1}{2}\mu|\nabla x - u^t + w^t|^2 \right\} \\ u^{t+1} &= \mathrm{argmin}_u \left\{ \tau \|u\|_1 + \tfrac{1}{2}\mu|\nabla x^{t+1} - u + w^t|^2 \right\} \\ w^{t+1} &= w^t + (\nabla x^{t+1} - u^{t+1}) \end{aligned} \quad (7)$$

Here, $t$ denotes the $t^{th}$ iteration. In the ADMM implementation procedure, these three subproblems are executed iteratively. Usually, it takes less than 30 iterations when the stopping criteria is reached, which cost about 0.025 seconds.

In the following, we conduct several simulations to demonstrate the reconstruction performance of the sample's reflection spectrum with our proposed method. We adopt a well-accepted evaluation criterion from computer vision to quantify the reconstruction accuracy. And the reconstruction accuracy is defined as

$$R_{dB} = 10 \log_{10} \left( \frac{\|r_c^{\mathrm{truth}}\|^2}{\|r_c^{\mathrm{truth}} - r_c^{\mathrm{recon}}\|^2} \right), \quad (8)$$

where $r_c^{\mathrm{truth}}$ denotes the truth value of the sample's reflection spectrum, and $r_c^{\mathrm{recon}}$ represents the reconstructed reflection spectrum. A large $R_{dB}$ corresponds to higher reconstruction accuracy.

The signal generation and processing flow is illustrated in Fig. 3. The generation process is as follows: we utilize a broadband light source in our simulations, as shown in Fig. 4. Here, the entire wavenumber range is divided into several wavenumber channels, and the marginal regions with extremely low intensities are omitted. The scanning interval is set to one-eighth of 0.57μm, and the sample surface height is randomly assigned. We calculate the complex reflection coefficients for different metallic materials, including gold (Au), silver (Ag), and copper (Cu), using the Fresnel equations. These conditions enable us to simulate a series of WLI fringe signals. Subsequently, we introduce 40 dB of Gaussian noise into the signals to mimic practical scenarios. For signal processing, we use Fourier domain analysis [4] to process either simulated signals or measured signals to estimate sample surface heights. With knowledge of the NA, spectral distribution of light source, scanning interval, and estimated surface height, a conditioned measurement matrix is created. The absolute reflection coefficient spectrum is then reconstructed by solving Eq. (6).

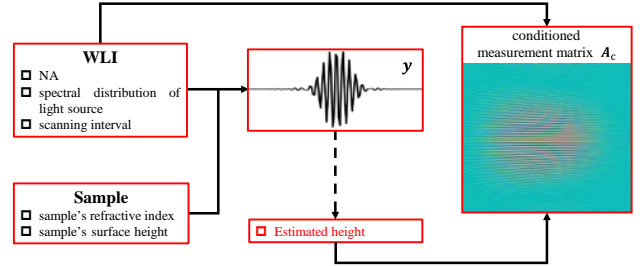

Fig. 3. Signal generation and processing flow for WLI signals.

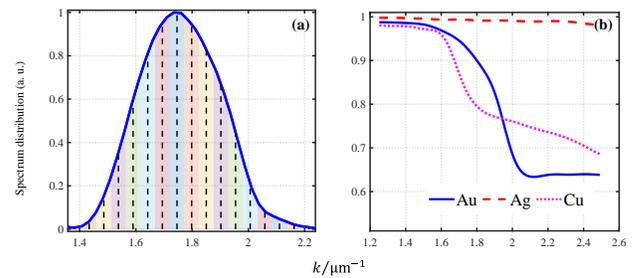

Fig. 4. Spectral distributions. (a) light source, (b) reflection coefficient spectra for Au, Ag, and Cu samples.

In Fig. 5, the simulated reconstruction performance of the reflectivity spectra-the square of the absolute reflection coefficient-is illustrated for various samples, including Ag, Au, and Cu, with comparisons to theoretical results at a 0.55 NA WLI. The average reconstruction accuracies are 18.8dB for Ag, 20.9dB for Au, and 22.7dB for Cu. Additionally, the standard deviation of the reconstruction accuracy varies across different wavenumber channels, particularly in those with low light source intensity. Meanwhile, A significant discrepancy is also observed between the reconstructed spectra and the theoretical values. Further explanations follow the illustrations of the experimental results.

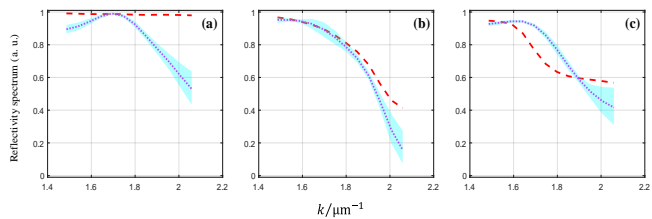

Fig. 5. Simulated reconstruction of the reflectivity spectra for various samples: (a) Ag, (b) Au, (c) Cu.

We collected WLI signals from a flat glass coated with various pure metallic materials including Ag, Au, and Cu, using a commercial instrument (Nexview NX2, ZYGO, USA). The measurement parameters were set to match those used in the simulations. The signal processing techniques employed are shown in Fig. 3.

In Fig. 6, experimental reconstruction of the reflectivity spectra is illustrated for the Ag, Au, and Cu samples. In the upper row of subfigures, the surface topography of these materials is shown. Using the sample surface heights information, we created the corresponding measurement matrix at each pixel, and obtained the reflectivity spectrum at each pixel. Since we are measuring homogeneous materials, the reconstructed spectra were averaged to obtain the mean value and standard deviation, as shown in the lower subfigures.

When comparing the results of the simulations and experiments, the reconstructed spectra for the Ag and Cu samples show similar trends. However, there is a noticeable discrepancy in the Au sample, which may be attributed to other factors.

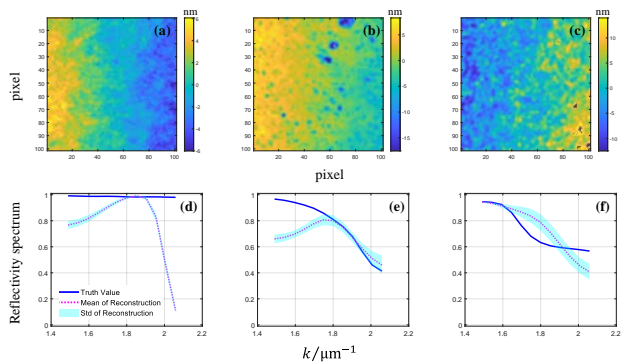

Fig. 6. Experimental reconstruction of the reflectivity spectra for various samples: (a) Ag, (b) Au, (c) Cu.

Before offering a conclusion, we would like to discuss the sources of error contributing to the discrepancy between the true values and the reconstructed spectra in both simulations and experiments. In the simulation, the dominant factor should be the phase change on reflection (PCOR) caused by the materials. This effect leads to phase misalignment in the monochromatic WLI signal at different wavenumbers, but it is not accounted for when constructing the measurement matrix. Although the estimated surface height slightly mitigates this effect, the influence of PCOR remains.

In the experimental case, additional errors arise from factors such as inaccurate NA information, non-uniform pupil field distribution, errors in scanning intervals, and chromatic dispersion of the optical system. These issues can introduce inaccuracies when creating the measurement matrix.

In conclusion, we propose a prior-enhanced reflection spectra imaging technique for high-NA WLI. This method leverages the inherent linear relationship between the WLI signal and the reflection spectra, incorporating matrix conditioning and regularization terms to achieve effective reconstruction of the material's reflectivity. While the reconstruction results in both simulations and experiments are not ideal, the mathematical framework opens possibilities for further improvement. Advanced signal processing techniques, such as AI or auto-differentiation, could be used to capture the PCOR of the measured sample. This could not only improve the reflectivity spectrum reconstruction but also enhance surface topography measurements.

**Funding.** National Natural Science Foundation of China (NSFC) (62105204, 52335010).

**Acknowledgement.** xxxxxxxxxx